\documentclass[prb,showpacs,twocolumn]{revtex4}
\usepackage{mathrsfs}
\usepackage{stmaryrd}
\usepackage{times}
\usepackage{amsmath}
\usepackage{amsfonts}
\usepackage{amssymb}
\usepackage{graphicx}
\usepackage{bm}
\begin{document}

\title{The electric current induced heat generation in a strongly interacting
quantum dot in the Coulomb blockade regime }
\author{Jie Liu$^{1}$ }
\author{Juntao Song$^{1}$ }
\author{Qing-feng Sun$^{1,\ast}$}
\author{X. C. Xie$^{2,1}$}
\affiliation{$^{1}$Beijing National Laboratory for Condensed Matter
Physics and Institute of Physics, Chinese Academy of Sciences,
Beijing 100190, China} \affiliation{$^{2}$Department of Physics,
Oklahoma State University, Stillwater, Oklahoma 74078, USA}

\begin{abstract}
The heat generation by an electric current flowing through a quantum
dot with the dot containing both electron-electron interaction and
electron-phonon interaction, is studied. Using the non-equilibrium
Keldysh Green's function method, the current-induced heat generation
is obtained. We find that for a small phonon frequency, heat
generation is proportional to the current. However, for a large
phonon frequency, heat generation is in general qualitatively
different from the current. It is non-monotonic with current and
many unique and interesting behaviors emerge. The heat generation
could be very large in the Coulomb blockade region, in which the
current is very small due to the Coulomb blockade effect. On the
other hand, in the resonant tunneling region, the heat generation is
very small despite a large current, an ideal condition for device
operation. In the curve of heat generation versus the bias, a
negative differential of the heat generation is exhibited, although
the current is always monotonously increasing with the bias voltage.

\end{abstract}

 \maketitle

\section{INTRODUCTION}
  With the continued miniaturization of
integrated circuits, quantum phenomena and new technological
problems are emerging quickly. An essential technological roadblock
of this miniaturization is the "power problem". With millions of
transistors assembled on a chip area no larger than a few square
centimeters, the chip-level power densities may reach to the order
of $100 W/cm^2$. The large amount of accumulated heat caused by the
current makes the chip's temperature rise to a level so high that
the chip may not function properly. There are two ways to reduce the
chip's temperature: one is to remove the heat as quickly as
possible, while the other is to suppress the heat generation. It is
unfortunate that the novel and complex device geometries normally
make heat removal more difficult and most of the new materials
appearing in device processing have lower thermal conductivities
than bulk silicon.\cite{pop} Thus, it is essential to uncover the
laws of heat generation induced by an electric current in nanoscale
devices and to investigate how the heat generation may be reduced.

The idea that electric current does work on conductors and causes
heating has been known since the mid-nineteenth century. It has been
well studied in macroscopic systems and is known as the effect of
joule heating; however, many open questions remain concerning
heating in nanoscale systems. For a net resistive nanoscale system,
the total heat dissipation is still equal to the input power $IV$,
with $I$ being the current $I$ and $V$ the bias voltage. In standard
transport theory, such as the Landauer-Buttiker formula or the
non-equilibrium Keldysh Green function (NEKGF) method, it is
normally assumed that the electron is elastically scattered in the
central region (called the scattering region) and that all heat
dissipation occurs in the reservoirs.\cite{haug} This assumption
does not provide a full answer to the dissipation problem. One can
see this from the following two aspects: (1) If the central region
contains the phonon degrees of freedom and the electron-phonon (e-p)
interaction, the inelastic scattering processes and heat dissipation
can occur in the central region. In this case, the simple assumption
of all dissipation occurring in the reservoirs is invalid. In fact,
in some recent experiments, it is found that the e-p interaction is
indeed strong in some nanoscale molecular devices, such as,
$\mbox{C}_{60}$, nanotube, etc.\cite{park,leroy,sapmaz} In these
devices, the phonon-assisted tunnelling sub-steps or sub-peaks have
been observed in the I-V curves or in the differential conductance
versus the gate voltage.\cite{park,leroy,sapmaz,chen,aref1} (2)
Furthermore, if one wishes to investigate where and how the heat
dissipation occurs while a current passes a nano-device even
assuming the device contains no e-p interaction, the e-p
interactions in the leads that near the nano-device must also be
considered. Then this part (i.e. the dissipation region) can not be
considered a part of the reservoirs since the reservoirs are assumed
to be without any interactions in the standard transport theory.

Recently, some attention has been paid to the problem of
current-induced heat generation in nanoscale devices, with a few
preliminary results.\cite{aref3,zhifeng,oron,lazzeri,aref4,a2ref1}
Huang \emph{et al.} have experimentally observed the current-induced
local heating effects in single molecules by measuring the force
required to break the molecule-electrode bonds.\cite{zhifeng} Very
recently, Oron-Carl and Krupke have successfully determined the hot
phonon generation from a current bias by using the ratio of
anti-Stokes to Stokes lines.\cite{oron} Theoretically, Lazzeri
\emph{et al.} have studied the heat generation in nanotube via the
first principles calculations.\cite{lazzeri} All previous results
demonstrate that the heat generation in nanoscale devices is
important. In a recent work, we studied the heat generation when an
electric current passes through a mesoscopic devices.\cite{sun} By
using the NEKGF method, a general formula for the current-induced
heat generation was derived. This formula can be applied to both the
linear and nonlinear transport region, as well the mesoscopic device
with various interactions.

In this paper, we study the current-induced heat generation while
an electric current flows through a QD that is in the Coulomb
blockade regime, focusing on the heat transfer to the phonon part
since the phonon creation is the main source of the heat
generation in QD. Considering that the device consists of a QD
coupled to two leads and the QD contains the electron-electron
(e-e) Coulomb interaction and e-p interaction, by using the NEKGF
method and the general formula of heat generation, both the heat
generation and the current are obtained. The numerical results
exhibit that the heat generation in the Coulomb blockade regime
has many interesting and unique behaviors. For example: (1) In the
resonant tunneling region with the intra-dot level within the bias
window, the current is very large but the current-induced heat
generation is very small. This is an ideal region for the
operation of the nano-device with a large current and low heat
generation. On the other hand, in the Coulomb blockade region, the
current is very small but the heat generation is very large. In
particular, when the e-e interaction strength is equal to the
phonon frequency, a resonant phonon emitting process occurs and a
large amount of current-induced heat generation emerges, even
though the current is quite small due to the Coulomb blockade
effect. (2) In some parameter regions, when the bias rises, the
current increases but the heat generation decreases. In other
words, the negative differential of the heat generation emerges.

\begin{figure}
\centering
\includegraphics[height=180pt,viewport=0 20 700 562,clip]{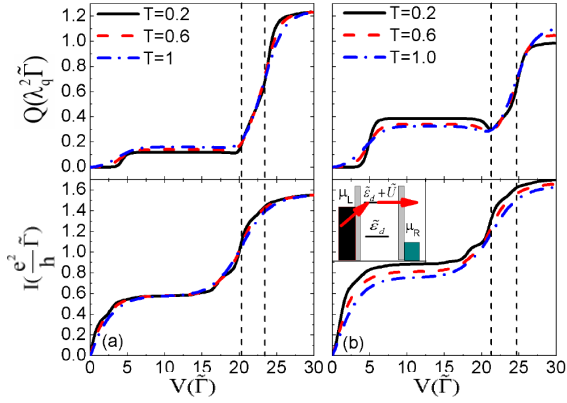}
\caption{(Color online) The heat generation $Q(\lambda_{q}^{2})$ and
current $I$ vs. the bias $V$ for the different temperature $T$ with
the level $\tilde{\epsilon}_d=0$ (a) and $1$ (b), and the other
parameter is $\omega_{q}=3,\tilde{U}=20,\lambda_{q}=0.6\omega_{q}$.
The inset in (b) is schematic diagram for the tunneling process in
which a phonon is absorbed. }\label{Q1}
\end{figure}
\begin{figure}
\centering
\includegraphics[height=140pt,viewport=0 2 725 571,clip]{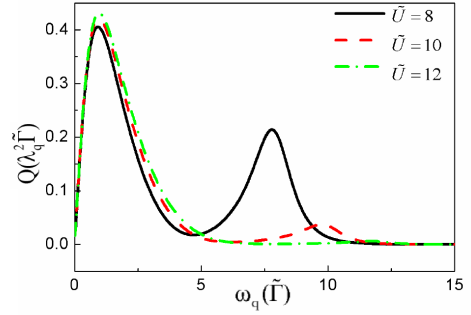}
\caption{ (Color online) The of heat generation $Q(\lambda_{q}^{2})$
vs. the phonon frequency $\omega_q$ for different e-e interaction
strength $\tilde{U}$ with the parameters: $V=4$, $T=1$,
$\lambda_{q}=0.6\omega_{q}$, and $\tilde{\epsilon}_d=0$}\label{Q2}
\end{figure}

\section{MODEL AND FORMULATION}
 The system of the lead-QD-lead can be
described by the following Hamiltonian:
\begin{eqnarray}\label{a}
 H & = &
 \sum\limits_{\alpha, k, \sigma  }\varepsilon_{\alpha k}\hat{c}_{\alpha k\sigma}^{\dag}\hat{c}_{\alpha
 k\sigma}
 + \sum\limits_{\alpha, k, \sigma}t_{\alpha
 k}(\hat{d}_{\sigma}^{\dag}\hat{c}_{\alpha k\sigma}+H.c.) \nonumber\\
 & + &
  \omega_{q}\hat{a}_{q}^{\dag}\hat{a}_{q}
 +\sum\limits_{\sigma}(\varepsilon_{d}+\lambda_{q}(\hat{a}_{q}^{\dag}+\hat{a}_{q}))
 \hat{n}_{\sigma} +U\hat{n}_{\uparrow}\hat{n}_{\downarrow},
\end{eqnarray}
where $\hat{n}_{\sigma}=
\hat{d}_{\sigma}^{\dagger}\hat{d}_{\sigma}$, and $\alpha =L, R$
represent the left and right leads. $\hat{c}_{\alpha
k\sigma}^{\dag}$ and $\hat{d}_{\sigma}^{\dagger}$ create an
electron with spin $\sigma$ in the $\alpha$ lead and QD,
respectively. Analogously, $\hat{a}_{q}^{\dag} $ is the phonon
creation operator and $\omega_{q}$ is the phonon frequency. The QD
has a single energy level $\varepsilon_{d}$ with the spin
$\sigma=\uparrow,\downarrow$. The intra-dot e-e Coulomb
interaction and e-p interaction are considered, with the
interaction strength $U$ and $\lambda_{q}$. The second term in Eq.
(\ref{a}) describes the tunneling coupling between the QD and the
two leads and $t_{\alpha k}$ is the hopping matrix element.

Before to solve the heat generation in the QD we first make a
canonical transform with the unitary operator\cite{mahan}
$\hat{U}=\exp\{-\sum_{\sigma}(\lambda_{q}/\omega_{q})
(\hat{a}^{\dag}_{q}-\hat{a}_{q})\hat{n}_{\sigma} \}$. After this
transformation, the Hamiltonian becomes:
\begin{align}\label{H2}
\tilde{H} = \sum\limits_\sigma  {\tilde \varepsilon _d \hat d^\dag
_\sigma \hat d_\sigma   + \tilde U\hat n_ \uparrow  } \hat n_
\downarrow   + \sum\limits_{\alpha k,\sigma } {\varepsilon _{\alpha
k } \hat c^\dag _{\alpha k\sigma } \hat c_{\alpha k\sigma } }
\nonumber\\+ \sum\limits_{\sigma ,\alpha, k } {(t_{\alpha k } \hat
c^\dag _{k\sigma } \hat d_\sigma  \hat X + H.c.)}  + \omega _q \hat
a_q^\dag  \hat a_q,
\end{align}
where $\tilde{\varepsilon}_{d}=\varepsilon
_d-\lambda_{q}^{2}/\omega_{q}$,
$\tilde{U}=U-2\lambda_{q}^2/\omega_{q}$, and $\hat{X} =
\exp\{-(\lambda_q/\omega_q)(\hat{a}_q^{\dagger}-\hat{a}_q)\}$. We
make an approximation to replace the operator $\hat{X}$ by its
mean value $\langle \hat{X}
\rangle=\exp\{-(\lambda_{q}/\omega_{q})^{2}(N_{q}+1/2)\}$,\cite{mahan}
where $N_{q}=1/[exp(\omega_q/k_BT)-1]$ refers to the phonon
number. This approximation was used in previous
studies,\cite{chen, mahan} and it is known to be valid when
$t_{\alpha k} \ll \lambda_q$. After taking this approximation the
phonon is decoupled with the electron. Then the formulate of the
current-induced heat generation $Q$ per unit time is:\cite{sun}
\begin{align}\label{bb}
 Q = Re\sum\limits_{\sigma } {\omega _q \lambda _q^2 \int {\frac{{d\omega }}{{2\pi }}
 \{ \tilde{G}_{\sigma \sigma }^ <  (\omega ) \tilde{G}_{\sigma \sigma }^ >  (\bar \omega )} }
  \nonumber\\- 2N_q [\tilde{G}_{\sigma \sigma }^ >  (\omega )\tilde{G}_{\sigma \sigma }^a (\bar \omega )
  + \tilde{G}_{\sigma \sigma }^r (\omega )\tilde{G}_{\sigma \sigma }^ >  (\bar \omega
  )]\}.
\end{align}
where $\bar{\omega}= \omega-\omega_q$, and
$\tilde{G}^{r,a,<,>}_{\sigma\sigma}(\omega)$ are the Fourier
transforms of $\tilde{G}^{r,a,<,>}_{\sigma\sigma}(t)$, which are the
standard QD's single-electron Green's functions of the Hamiltonian
(\ref{H2}) and are defined as
$\tilde{G}^{r}_{\sigma\sigma}(t)=-i\Theta(t)\langle
\{d(t),d^{\dag}(0)\} \rangle,\,
\tilde{G}^{a}_{\sigma\sigma}(t)=i\Theta(-t)\langle\{d(t),d^{\dag}(0)\}\rangle,\,
\tilde{G}^{<}_{\sigma\sigma}(t)=i\langle d^{\dag}(0)d(t)\rangle,\,
\tilde{G}^{>}_{\sigma\sigma}(t)=-i\langle d(t)d^{\dag}(0)\rangle$.
Here heat generation $Q$ is the dissipation in the QD region in
which the energy is transferred to the local phonons. Since the
present device is a net resistive system, the total dissipation is
equal to the input power $IV$, and remaining dissipation $Q_{res}$
occurs in the reservoirs.\cite{note} Moreover the heat generation
actually is the rate of heat generation. Hereafter we don't
distinguish them because that they have the completely same
characteristic. Because of the e-p interaction has been decoupled in
the Hamiltonian (\ref{H2}), the Green's function
$\tilde{G}^{r}_{\sigma\sigma}$ can be obtained by the
equation-of-motion technique:\cite{meir,haug}
\begin{eqnarray}\label{c}
\tilde{G}^{r}_{\sigma\sigma}(\omega) & = &
\frac{[\omega_{\varepsilon}-\tilde{U}(1- \langle
n_{\bar{\sigma}}\rangle)]}
{\omega_{\varepsilon}(\omega_{\varepsilon}-\tilde{U})
-i\tilde{\Gamma}[\omega_{\varepsilon}-\tilde{U}(1- \langle
n_{\bar{\sigma}}\rangle )]},
\end{eqnarray}
where $\omega_{\varepsilon}\equiv \omega-\tilde{\varepsilon}_{d}$,
$\langle n_{\bar\sigma} \rangle $ is the electron occupation number
in the intra-dot state $\bar\sigma$, $\bar\sigma = \downarrow$ while
$\sigma =\uparrow$ and $\bar\sigma = \uparrow$ while $\sigma
=\downarrow$, $\tilde{\Gamma} = (\tilde{\Gamma}_L
+\tilde{\Gamma}_R)/2$, $\tilde{\Gamma}_{\alpha} = \Gamma_{\alpha}
\langle X\rangle^2$, and $\Gamma_{\alpha} =\sum_k 2\pi |t_{\alpha
k}|^2 \delta(\omega-\epsilon_{\alpha k})$ is the linewidth functions
which assume to be independent of the energy
$\omega$.\cite{haug,meir} After solving
$\tilde{G}^{r}_{\sigma\sigma}(\omega)$,
$\tilde{G}^{a}_{\sigma\sigma}(\omega) =
[\tilde{G}^{r}_{\sigma\sigma}(\omega)]^*$,
$\tilde{G}^{<}_{\sigma\sigma}(\omega) =
\tilde{G}^{r}(\omega)\tilde{\Sigma}^{<}(\omega)\tilde{G}^{a}(\omega)
$, and $\tilde{G}^{>}_{\sigma\sigma}(\omega) =
\tilde{G}^{<}_{\sigma\sigma} +\tilde{G}^{r}_{\sigma\sigma}
-\tilde{G}^{a}_{\sigma\sigma}$, where $\tilde{\Sigma}^<(\omega) =
i[\tilde{\Gamma}_L f_L(\omega) + \tilde{\Gamma}_R f_R(\omega)]$ and
$f_{\alpha}(\omega)= 1/\{exp[(\omega-\mu_{\alpha})/k_B T] +1\}$ is
the Fermi distribution function in the $\alpha$ lead. Substituting
these Green's functions into the Eq.(\ref{bb}), the heat generation
$Q$ can be obtained straightforwardly. Finally, the electron
occupation numbers $\langle n_{\sigma} \rangle $ in Eq.(\ref{c})
need to be self-consistently calculated with the equation $\langle
n_{\sigma} \rangle  = i \int (d\omega/2\pi)
\tilde{G}^<_{\sigma\sigma}(\omega)$. In addition, by using the same
method as in Ref.\cite{chen}, the current can also be calculated.
\begin{figure}
\centering
\includegraphics[height=300pt,viewport=0 0 556 760,clip]{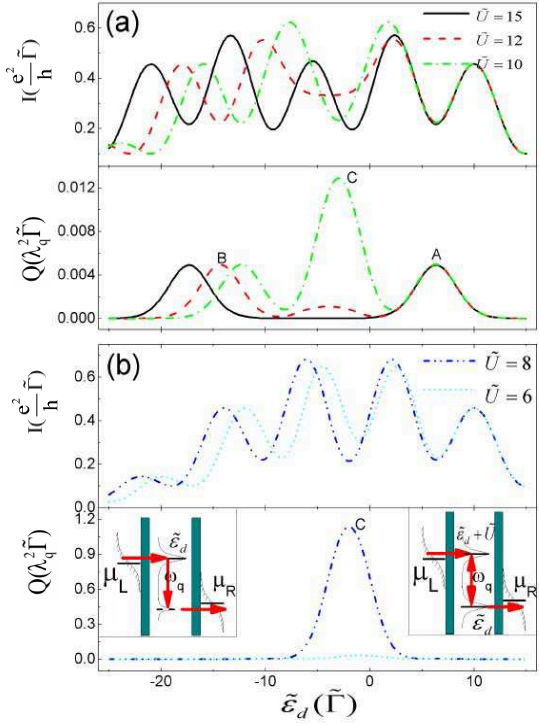}
\caption{ (Color online)  $Q(\lambda_{q}^{2})$ and $I$ vs. the level
$\tilde{\epsilon}_d$ for the different $\tilde{U}$ with the
parameters $\omega_{q}=8$, $V=4$, $T=1$, and
$\lambda_{q}=0.6\omega_{q}$. The insets in (b) are the schematic
plots for two phonon-emitting processes. }\label{Q3}
\end{figure}

\section{NUMERICAL RESULT}
 In the numerical investigation, we consider
the symmetric barriers\cite{addnote2} with $\tilde{\Gamma}_L=
\tilde{\Gamma}_R =\tilde{\Gamma}$, and set $\tilde{\Gamma}=1$ as the
unit of energy and $\mu_R=0$ as the energy zero point, then the bias
$eV= \mu_L-\mu_R =\mu_L$. Fig.1a shows the heat generation $Q$ and
current $I$ as functions of the bias $V$ for the different
temperature $T$ at the QD level $\tilde{\epsilon}_d =0$. We can see
that at first the heat generation $Q$ is almost zero while the bias
$V<\omega_q$ since the bias can not offer enough energy to emit in
this range. With the bias passing $\omega_q$, both of $Q$ and $I$
rapidly increase and reach their respective plateaus. $Q$ and $I$
remain at plateaus until the bias reaches $\tilde{U}$, then they
rapidly increase and reach new plateaus. The reason behind the two
steps is that the QD's level is split into two sub-levels at
$\tilde{\epsilon}_d$ and $\tilde{\epsilon}_d +\tilde{U}$ with the
existence of the e-e interaction $\tilde{U}$. The temperature has
little effect on the heat generation. Raising temperature makes the
curves of both $Q$-$V$ and $I$-$V$ more smooth, but the qualitative
behaviors remain. Here we emphasize that the heat generation $Q$ is
not directly proportional to the current $I$, and they obviously has
the following two differences: (i) Many small sub-steps emerge in
the current curves because of the phonon-assisted tunneling
processes, but no phonon-assisted sub-steps in the heat generation
curves. (ii) The rapid jumps in the current curve happen at about
$V=\tilde{\epsilon}_d$ and $\tilde{\epsilon}_d +\tilde{U}$, but the
rapid jumps in the heat generation curve happen at about
$V=\tilde{\epsilon}_d +\omega_q$ and $\tilde{\epsilon}_d +\tilde{U}
+\omega_q$, a delay of $\omega_q$ (see Fig.1a).

In Fig.1a, the level $\tilde{\epsilon}_d$ is fixed at zero. Next, we
study the case of $\tilde{\epsilon}_d\not=0$. Fig.1b describes the
heat generation $Q$ with the same parameters as in Fig.1a except
$\tilde{\epsilon}_d=1$. The results clearly exhibit that a negative
differential heat generation $dQ/dV$ emerges when $V$ is between
$\tilde{U}+\tilde{\epsilon}_d-\omega_q$ and
$\tilde{U}+\tilde{\epsilon}_d$. In this region, with increasing $V$,
$I$ increases, but $Q$ decreases. The reason is that the electron
can absorb as well as emit a phonon while it tunnels through the QD.
When the bias $V$ is between $\tilde{U}+\tilde{\epsilon}_d-\omega_q$
and $\tilde{U}+\tilde{\epsilon}_d$, the tunneling process as shown
in the inset of Fig.1b occurs, in which the electron absorbs a
phonon while it passes through the QD. Thus, a negative differential
heat generation emerges.

Fig.2 investigates the dependence of the heat generation $Q$ on
the phonon frequency $\omega_q$. When $\omega_q$ approaches to
zero, so does the heat generation $Q$. With $\omega_q$ increasing
from zero, $Q$ quickly increases. The curve of $Q$-$\omega_q$
exhibits two peaks. One peak is roughly at
$\omega_q=\tilde{\Gamma}$, very close to zero. The other peak is
at $\omega_q=\tilde{U}$. The first peak is caused by competition
of two effects when raising $\omega_q$: (1) the electron losses
more energy when it emits a phonon, at the same time (2) it is
more and more difficult to emit a phonon. The second peak is
largely associated with the e-e interaction strength $\tilde{U}$.
Its location is fixed at $\omega_q=\tilde{U}$ and its height
strongly depends on $\tilde{U}$. The smaller $\tilde{U}$ is, the
higher the peak is. In fact, this peak origins from a resonant
phonon emitting process as shown in the right inset of Fig.3b,
which will be discussed later in more detail.

Fig.3 shows $Q$ and $I$ versus the QD's level $\tilde{\epsilon}_d$
for different values of the e-e interaction $\tilde{U}$. $I$ has two
main resonant peaks at $\tilde{\epsilon}_d =\bar{\mu}$ and
$\tilde{\epsilon}_d +\tilde{U}=\bar{\mu}$, with $\bar{\mu}\equiv
(\mu_L+\mu_R)/2$. In addition, some phonon-assisted sub-peaks also
emerge at the two side of the main peaks, and the distance to its
main peak is $\pm n \omega_q$ ($n=1,2,3,...$). These well-known
results on current have been experimentally
observed\cite{park,leroy,sapmaz}. Following, we focus on the heat
generation $Q$. The curves of $Q$ versus the level
$\tilde{\epsilon}_d$ have three peaks, which are marked 'A', 'B',
and 'C'. The peak 'A' ('B') stands at the valley between the right
(left) main peak and its right-side (left-side) sub-peak of the
current. The peak 'C' places at $\tilde{\epsilon}_d +\tilde{U}/2
=\bar{\mu}$, i.e. at the valley (the Coulomb blockade region)
between the two main peaks of the current. Here we emphasize that
none of the peaks of $Q$ aligns with that of $I$, in fact all peaks
of $Q$ appear at the valleys of $I$. This means that the heat
generation is small for large current, and vice verse. Here we also
emphasize another important feature: with the e-e interaction
$\tilde{U}$ approaching to the phonon frequency $\omega_q$, the peak
'C' is greatly enhanced. At $\tilde{U} =\omega_q$, the peak 'C' is
very large. For example, the peak 'C' in Fig.3b with $\tilde{U}
=\omega_q =8$ is hundredfold higher than the peak 'C' in Fig.3a with
$\tilde{U}=10$. Due to the largeness of peak 'C', the peaks 'A' and
'B' are not visible in Fig.3b, although they are there. As far as
for the current, there is no such high peak at all.

Let us explain why the heat generation $Q$ is large at the valleys
of the current. This is because the phonon-emitting processes are
stronger in the current valley region. For example, to consider the
phonon-emitting tunneling process in the left inset of Fig.3b, in
which an electron from the left lead tunnels into the QD at the
level $\tilde{\epsilon}_d$, emits a phonon and jumps to the
sub-level $\tilde{\epsilon}_d -\omega_q$ due to the e-p interaction,
then tunnels to the right lead. The strength of this process is
direct ratio of the occupation probability $f_L(\tilde{\epsilon}_d)$
of the left lead at the energy $\omega=\tilde{\epsilon}_d$, and the
empty probability $1-f_R(\tilde{\epsilon}_d -\omega_q)$ of the right
lead at the energy $\omega=\tilde{\epsilon}_d -\omega_q$. So this
process is maximum at $\tilde{\epsilon}_d -\omega_q/2 =\bar{\mu}$,
and leads to the peak 'A' in the curve of $Q$-$\tilde{\epsilon}_d$.
The origin of the peak 'B' is same as for the peak 'A'. The peak 'C'
origins from the tunneling process as shown in the right inset of
Fig.3b, which is similar to that in the left inset of Fig.3b. But
the two levels $\tilde{\epsilon}_d$ and $\tilde{\epsilon}_d
+\tilde{U}$ are real electronic levels now. Therefore the resonant
phonon emitting occurs when the difference ($\tilde{U}$) of two
levels is equal to the phonon frequency $\omega_q$, giving rise to a
huge peak in heat generation.

\begin{figure}
 \centering
\includegraphics[height=120pt,viewport=16 32 705 424,clip]{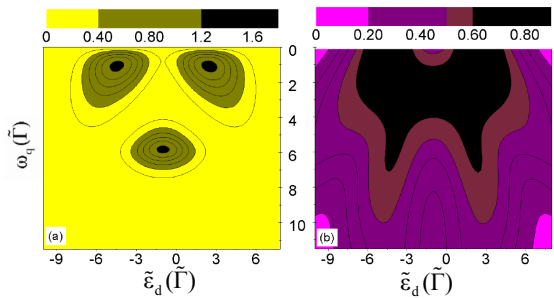}
\caption{$Q(\lambda_{q}^{2})$ (a) and $I$ (b) vs. the level
$\tilde{\epsilon}_d$ and phonon frequency $\omega_q$ with
$\tilde{U}=6$, $V=4$, $T=1$, and
$\lambda_{q}=0.6\omega_{q}$}\label{Q6}
\end{figure}

Finally, in Fig.4 we present the contour plot of the heat generation
$Q$ and current $I$ versus the QD level $\tilde{\epsilon}_d$ and
phonon frequency $\omega_q$. As $\omega_q$ approaching to zero, $Q$
also approaches to zero regardless of the other parameters, however,
$I$ can be large if the resonant tunneling occurs. At small but
non-zero $\omega_q$, e.g. $\omega_q <\tilde{\Gamma}$, the heat
generation $Q$ is approximatively proportional to the current $I$.
In this case the quantum result of the heat generation resembles the
classical one. On the other hand, at large $\omega_q$ (e.g.
$\omega_q>\tilde{\Gamma}$), the contour of the heat generation is
totally different to the contour of the current. For example, for
$\omega_q=\tilde{U}$, a great heat generation appears, but we can
not find any large change in the current plot.

\section{CONCLUSION}
In summary, we investigated the heat generation in a quantum dot
(QD) under a bias and in the Coulomb blockade regime. The results
show that for the small phonon frequency, the heat generation is
proportional to the current. On the other hand, for the large phonon
frequency, the heat generation is quite different from the current
and exhibits many unique and interesting characteristics. For
instance, a negative differential heat generation emerges in some
parameter regions even though the differential conductance is always
positive. In particular, the heat generation is large in the Coulomb
blockade region, while it is small in the resonant tunneling region.
The later behavior is advantageous for the QD device because the QD
device normally works in the resonant tunneling region and low heat
generation is ideal for device operation.

\section*{ACKNOWLEDGE}
 We gratefully acknowledge the financial support
from the Chinese Academy of Sciences, NSF-China under Grants Nos.
10525418, 10734110, and 10821403 and National Basic Research
Program of China (973 Program project No. 2009CB929100). X.C.X. is
supported by US-DOE under Grants No. DE-FG02- 04ER46124.


\begin{thebibliography}{99}
\bibitem{note}
Notice that $Q$ can be larger than $IV$ although $Q+Q_{res}=IV$. For
example, in the Fig.3b $Q\approx 1.14 \lambda_q^2$ at the highest
point of the peak 'C', which is greatly larger than $IV\approx
0.215\times 4 $.
\bibitem{addnote2}
For the nonsymmetric barriers, the heat generation $Q$ is directly
proportion to the value in the symmetric case, thus, all results
remain the same.
\end{thebibliography}
\end{document}